\documentclass[english,aps]{article}
\usepackage[T1]{fontenc}
\usepackage[latin1]{inputenc}
\usepackage{graphicx}
\usepackage{amssymb}
\usepackage{sublabel}
\usepackage{simplemargins}

\setallmargins{1in}
\makeatletter

\newcommand{\comment}[1]{}

\usepackage{epsfig}

\usepackage{babel}
\makeatother
\begin{document}

\title{Full solution for the storage of correlated memories in an autoassociative memory}
\author{Emilio Kropff\footnote{kropff@sissa.it - http://people.sissa.it/$\sim$kropff} \\\\
 SISSA - International School of Advanced Studies\\
 via Beirut 4\\
34014, Trieste\\
 Italy}

\maketitle
\begin{abstract}
We complement our previous work \cite{hfsp} with the full (non diluted) solution describing the stable states of an attractor network that stores correlated patterns of activity. The new solution provides a good fit of simulations of a network storing the feature norms of McRae and colleagues \cite{McRae_norms}, experimentally obtained combinations of features representing concepts in semantic memory. We discuss three ways to improve the storage capacity of the network: adding uninformative neurons, removing informative neurons and introducing popularity-modulated hebbian learning. We show that if the strength of synapses is modulated by an exponential decay of the \textit{popularity} of the pre-synaptic neuron, any distribution of patterns can be stored and retrieved with approximately an optimal storage capacity - i.e, $C_{min}\propto I_f p$, the minimum number of connections per neuron needed to sustain the retrieval of a pattern is proportional to the information content of the pattern multiplied by the number of patterns stored in the network.
\end{abstract}

\section{Introduction}
Autoassociative memory networks can store patterns of neural activity by modifying the synaptic weights that inter-connect neurons \cite{Hopfield, Amit_1989}, following the Hebbian rule \cite{Hebb}. Once a pattern of activity is stored, it becomes an attractor of the dynamics of the system. Direct evidence showing attractor behavior in the hippocampus of \textit{in vivo} animals has been reported \cite{OKeefe_2005}. These kind of memory systems have been proposed to be present at all levels along the cortex of higher order brains, where hebbian plasticity plays a major role. 

Most models of autoassociative memory studied in literature store patterns that are obtained from some random distribution. Some exceptions appeared during the 80's when interest grew around the storage of patterns derived from hierarchical trees \cite{Parga_1986, Gutfreund_1988}. Of particular interest, Virasoro \cite{Virasoro_1988} relates the behavior of networks of general architecture with $prosopagnosia$, an impairment that impedes a patient to individuate certain stimuli without affecting its capacity to categorize them. Interestingly, the results from this model indicate that prosopagnosia is not present in Hebbian-plasticity derived networks. Some other developments have used perceptron-like or other arbitrary local rules for storing generally correlated patterns \cite{Gardner_1989, Diederich_1987} or patterns with spatial correlation \cite{Monasson_1992}. More recently, Tsodyks and collaborators \cite{Tsodyks_2006} have studied a Hopfield memory in which a sequence of morphs between two uncorrelated patterns are stored. In this work, the use of a saliency function favouring unexpected over expected patterns during learning results in the formation of a continuous one-dimensional attractor that spans the space between the two original memories. The fusion of basins of attraction can be an interesting phenomenon that we are not going to treat in this work, since we assume that the elements stored in a memory such as the semantic one are differentiable by construction.

Feature norms are a way to get an insight on how semantic information is organized in the human brain \cite{Vigliocco_2002, Garrard_2001, McRae_norms}. The information is collected by asking different types of questions about particular concepts to a large population of subjects. Representations of the concepts are obtained in terms of the features that appear more often in the subjects' descriptions. In this work we analyze the feature norms of McRae and colleagues \cite{McRae_norms} for two reasons: they are public and the size of the dataset allows a statistical approach (it includes $541$ concepts described in terms of $2526$ features). The norms were downloaded from the \textit{Psychonomic Society Archive of Norms, Stimuli, and Data} web site (www.psychonomic.org/archive) with consent of the authors.

In section \ref{model} we define a simple binary associative network, showing how it can be modified in order to store correlated representations. In section \ref{self} we solve the equilibrium equations for the stable attractor states of the system using a self-consistent signal to noise approach. Finally, in section \ref{norms} we study the storage of the feature norms of McRae and colleagues representing semantic memory elements. 

\section{The model}
\label{model}

We assume a network with $N$ neurons and $C\leq N$ synaptic connections per neuron. If the network stores $p$ patterns, the parameter $\alpha = p/C$ is a measure of the memory load normalized by the size of the network. In classical models, the equilibrium properties of large enough networks depends on $p$, $C$ and $N$ only through $\alpha$, which allows the definition of the thermodynamic limit ($p\rightarrow \infty$, $C\rightarrow \infty$, $N\rightarrow \infty$, $\alpha$ constant).

The activity of neuron $i$ is described by the variable $\sigma_i$, with $i=1...N$. Each of the $p$ patterns is a particular state of activation of the network. The activity of neuron $i$ in pattern $\mu$ is described by $\xi_i^{\mu}$, with $\mu=1...p$. The perfect retrieval of pattern $\mu$ is thus characterized by $\sigma_i=\xi_i^{\mu}$ for all $i$. We will assume binary patterns, where $\xi_i^{\mu}=0$ if the neuron is silent and $\xi_i^{\mu}=1$ if the neuron fires. Consistently, the activity states of neurons will be limited by $0\leq \sigma_i\leq 1$. We will further assume a fraction $a$ of the neurons being activated in each pattern. This quantity receives the name of $sparseness$.

Each neuron receives $C$ synaptic inputs. To describe the architecture of connections we use a random matrix with elements $C_{ij}=1$ if a synaptic connection between post-synaptic neuron $i$ and pre-synaptic neuron $j$ exists and $C_{ij}=0$ otherwise, with $C_{ii}=0$ for all $i$. In addition to this, synapses have associated weights $J_{ij}$. 

The influence of the network activity on a given neuron $i$ is represented by the field

\begin{equation}
h_i=\sum_{j=1}^N C_{ij}J_{ij}\sigma_j \label{eq:field}
\end{equation}
which enters a sigmoidal activation function in order to update the activity of the neuron

\begin{equation}
\sigma_i=\left\{1+\exp \beta \left(U-h_i\right)\right\}^{-1}\label{eq:activation}
\end{equation}
where $\beta$ is inverse to a temperature parameter and $U$ is a threshold favoring silence among neurons \cite{Buhmann_1989, Tsodyks_1988}.

The learning rule that defines the weights $J_{ij}$ must reflect the Hebbian principle: every pattern in which both neurons $i$ and $j$ are active will contribute positively to $J_{ij}$. In addition to this, the rule must include, in order to be optimal, some prior information about pattern statistics. In a one-shot learning paradigm, the optimal rule uses the sparseness $a$ as a learning threshold,

\begin{equation}
J_{ij}=\frac{1}{C a}\sum_{\mu=1}^p\left(\xi_i^{\mu}-a\right)\left(\xi_j^{\mu}-a\right).\label{eq:oneshot}
\end{equation}

However, as we have shown in previous work \cite{hfsp}, in order to store correlated patterns this rule must be modified using $a_j$, or the \textit{popularity} of the pre-synaptic neuron, as a learning threshold,

\begin{equation}
J_{ij}=\frac{1}{C a}\sum_{\mu=1}^p \xi_i^{\mu}\left(\xi_j^{\mu}-a_j\right) \label{eq:newrule},
\end{equation}
with
\begin{equation}
a_i\equiv \frac{1}{p}\sum_{\mu=1}^p \xi_i^{\mu}\label{eq:sharedness}.
\end{equation}
This requirement comes from splitting the field into a signal and a noise part,
\begin{equation}
h_i=\frac{1}{C a}\xi_i^{1}\sum_{j=1}^N C_{ij}\left(\xi_j^{1}-a_j\right)\sigma_j + \frac{1}{C a}\sum_{\mu=2}^p\xi_i^{\mu}\sum_{j=1}^N C_{ij}\left(\xi_j^{\mu}-a_j\right)\sigma_j\nonumber,
\end{equation}
and, under the hypothesis of gaussian noise, setting the average to zero and minimizing the variance. This last is
\begin{eqnarray}
var = & &\frac{1}{C^2 a^2}\sum_{\mu=1}^p\xi_i^{\mu}\sum_{j=1}^N C_{ij}\sigma_j^2\left(\xi_j^{\mu} -a_j\right)^2+   \nonumber\\
& + &\frac{1}{C^2 a^2}\sum_{\mu\neq\nu=1}^p\xi_i^{\mu}\xi_i^{\nu}\sum_{j=1}^N C_{ij}\sigma_j^2\left(\xi_j^{\mu} -a_j\right)\left(\xi_j^{\nu} -a_j\right)+\nonumber\\
& + &\frac{1}{C^2 a^2}\sum_{\mu=1}^p\xi_i^{\mu}\sum_{j\neq k=1}^N C_{ij}C_{ik}\sigma_j \sigma_k\left(\xi_j^{\mu} -a_j\right)\left(\xi_k^{\mu} -a_k\right)+\nonumber\\
& + &\frac{1}{C^2 a^2}\sum_{\mu\neq\nu=1}^p\xi_i^{\mu}\xi_i^{\nu}\sum_{j\neq k=1}^N C_{ij}C_{ik}\sigma_j \sigma_k\left(\xi_j^{\mu} -a_j\right)\left(\xi_k^{\nu} -a_k\right).
\label{eq:variance4}
\end{eqnarray}
If statistical independence is granted between any two neurons, only the first term in Eq. \ref{eq:variance4} survives when averaging over $\{\xi\}$.

\begin{figure}[ht]
\centerline{\hbox{\epsfig{figure=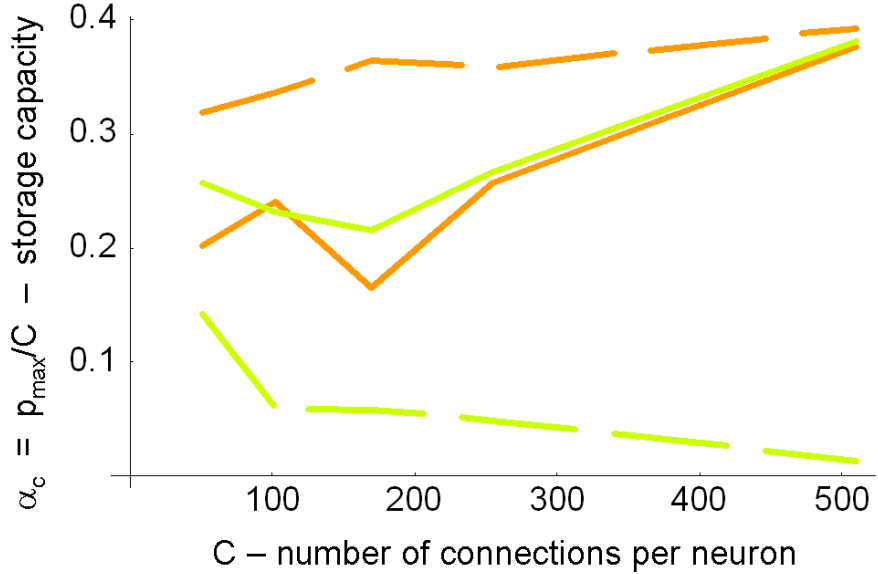,width=12cm,angle=0}}}
\caption{The four combinations of two learning rules and two types of dataset. Green: one shot 'standard' learning rule of Eq. \ref{eq:oneshot}. Orange: modified rule of Eq. \ref{eq:newrule}. Solid: trivial distribution of randomly correlated patterns obtained from Eq. \ref{eq:trivial}. Dashed: non-trivially correlated patterns obtained using a hierarchical algorithm. In three cases, the storage capacity (the maximum number of retrievable patterns normalized by $C$) with $C$ (the number of connections per neuron) is finite and converges to a common value as $C$ increases. Only in the case of one-shot learning of correlated patterns there is a storage collapse.}
\label{fig:scaling}
\end{figure}

In Figure \ref{fig:scaling} we show that the rule in Eq. \ref{eq:oneshot} can effectively store uncorrelated patterns taken from the distribution
\begin{equation}
P\left(\xi_i^{\mu}\right)=a\delta\left(\xi_i^{\mu}-1\right)+\left(1-a\right)\delta\left(\xi_i^{\mu}\right). \label{eq:trivial}
\end{equation}
but cannot handle less trivial distributions of patterns, suffering a storage collapse. The storage capacity can be brought back to normal by using the learning rule in Eq. \ref{eq:newrule}, which is also suitable for storing uncorrelated patterns.

Having defined the optimal model for the storage of correlated memories, we analyze in the following sections the storage properties and its consequences through mean field equations.

\section{Self consistent analysis for the stability of retrieval}
\label{self}

We now proceed to derive the equations for the stability of retrieval, similarly to what we have done in \cite{hfsp} but in a network with an arbitrary level of random connectivity, where the approximation $C\ll N$ is no longer valid \cite{Shiino_1992,Shiino_1993,yasser_2004}. Furthermore, we introduce patterns with variable mean activation, given by 

\begin{equation}
d_{\mu}\equiv \frac{1}{N} \sum_{j=1}^N{\xi_j^{\mu}}
\end{equation}
for a generic pattern $\mu$. As a result of this, the optimal weights are given by

\begin{equation}
J_{ij}= g_j \sum_{\mu=1}^p \frac{c_{ij}}{C d_{\mu}}\xi_i^{\mu}(\xi_j^{\mu}-a_j)
\end{equation}
which ensures that patterns with different overall activity will have not only a similar noise but also a similar signal. In addition, we have introduced a factor $g_j=g(a_j)$ in the weights that may depend on the popularity of the pre-synaptic neuron. We will consider $g_j=1$ for all but the last section of this work.

If the generic pattern $1$ is being retrieved, the field in Eq. \ref{eq:field} for neuron $i$ can be written as a signal and a noise contribution

\begin{equation}
h_i=\xi_i^{1} m_i^{1}+\sum_{\mu \neq 1} \xi_i^{\mu} m_i^{\mu}\label{eq:sigto}
\end{equation}
with
\begin{equation}
m_i^{\mu}=\frac{1}{C d_{\mu}} \sum_{j=1}^N g_j c_{ij} (\xi_j^{\mu}-a_j) \sigma_j \label{eq:mi}.
\end{equation}
We hypothesize that in a stable situation the second term in Eq.\ref{eq:sigto}, the noise, can be decomposed into two contributions
\begin{equation}
\sum_{\mu \neq 1} \xi_i^{\mu} m_i^{\mu}=\gamma_i \sigma_i+\rho_i z_i.\label{eq:self}
\end{equation}
The second term in Eq. \ref{eq:self} represents a gaussian noise with standard deviation $\rho_i$, and $z_i$ a random variable taken from a normal distribution of unitary standard deviation. The first term is proportional to the activity of the neuron $i$ and results from closed synaptic loops that propagate this activity through the network back to the original neuron, as shown in \cite{yasser_2004}. As is typical in the self consistent method, we will proceed to estimate $m_i^{\mu}$ from the ansatz in Eq. \ref{eq:self}, inserting it into Eq. \ref{eq:sigto} and validating the result with, again, Eq. \ref{eq:self}, checking the consistency of the ansatz.

Since Eq. \ref{eq:self} is a sum of $p\rightarrow \infty $ microscopic terms, we can take a single term $\nu$ out and assume that the sum changes only to a negligible extent. In this way, the field becomes
\begin{equation}
h_i\simeq \xi_i^1 m_i^1+\xi_i^{\nu} m_i^{\nu}+\gamma_i \sigma_i+\rho_i z_i.
\end{equation}
If the network has reached stability, which we assume, updating neuron $i$ does not affect its state. This can be expressed by inserting the field into Eq. \ref{eq:activation}, 
\begin{equation}
\sigma_i=\left\{1+\exp(-\beta(h_i-U))\right\}^{-1}\equiv G \left[\xi_i^1 m_i^1+\xi_i^{\nu} m_i^{\nu}+\rho_i z_i \right]. \label{eq:g}
\end{equation}

In the RHS of Eq. \ref{eq:g} the contribution of $\gamma_i \sigma_i$ to the field has been reabsorbed into the definition of $G[x]$. At first order in $\xi_j^{\nu} m_j^{\nu}$, Eq. \ref{eq:g} corresponding to neuron $j$ can be written as

\begin{equation}
\sigma_j\simeq G\left[\xi_j^1 m_j^1+\rho_j z_j \right]+G'\left[\xi_j^1 m_j+\rho_j z_j \right]\xi_j^{\nu} m_j^{\nu}. \label{eq:aprox}
\end{equation}
To simplify the notation we will further use $G_j\equiv G\left[\xi_j^1 m_j^1+\rho_j z_j \right]$ and $G_j'\equiv G'\left[\xi_j^1 m_j+\rho_j z_j \right]$. To this order of approximation, Eq. \ref{eq:mi} becomes
\begin{equation}
m_i^{\mu}= \frac{1}{C d_{\mu}}\sum_{j=1}{N}g_j c_{ij}(\xi_j^{\mu}-a_j) \left\{ G_j +G_j' \xi_j^{\mu} m_j^{\mu}\right\}\label{eq:maprox}.
\end{equation}
Other terms of the same order in the Taylor expansion could have been introduced in Eq. \ref{eq:aprox}, corresponding to the derivatives of $G$ with respect to $\xi_j^{\mu}m_j^{\mu}$ for $\mu\neq\nu$. It is possible to show, however, that such terms give a negligible contribution to the field.

If we define
\begin{eqnarray}
L_i^{\mu}&=&\frac{1}{C d_{\mu}} \sum_{j=1}^N g_j c_{ij} (\xi_j^{\mu}-a_j) G_j\nonumber\\
K_{ij}^{\mu}&=&\frac{1}{C d_{\mu}} g_j c_{ij}(\xi_j^{\mu}-a_j)\xi_j^{\mu}G_j'\label{eq:lk},
\end{eqnarray}
Eq. \ref{eq:maprox} can be simply expressed as
\begin{equation}
m_i^{\mu}=L_i^{\mu}+\sum_{j=1}^N K_{ij}^{\mu}m_j^{\mu}\label{eq:recurrent}.
\end{equation}
This equation can be applied recurrently to itself renaming indexes, 
\begin{equation}
m_i^{\mu}=L_i^{\mu}+\sum_{j=1}^N K_{ij}^{\mu}L_j^{\mu}+\sum_{j=1}^N\sum_{k=1}^N K_{ij}^{\mu} K_{jk}^{\mu}m_k^{\mu}.
\end{equation}
If applied recurrently infinite times, this procedure results in
\begin{equation}
m_i^{\mu}=L_i^{\mu}+\sum_{j=1}^N K_{ij}^{\mu}L_j^{\mu}+\sum_{j=1}^N\sum_{k=1}^N K_{ij}^{\mu} K_{jk}^{\mu}L_k^{\mu}+\dots
\end{equation}
which, by exchanging mute variables, can be re-written as
\begin{equation}
m_i^{\mu}=L_i^{\mu}+\sum_{j=1}^N L_j^{\mu} \left\{ K_{ij}^{\mu} +\sum_{k=1}^N K_{ik}^{\mu}K_{kj}^{\mu}+\sum_{k,l=1}^N K_{ik}^{\mu}K_{kl}^{\mu}K_{lkj}^{\mu}+\dots\right\}.\label{eq:infinite}
\end{equation}

Eq. \ref{eq:infinite} can be decomposed into the contribution of the activity of $G_i$ on one side and that of the rest of the neurons on the other, which will correspond to the first and the second term on the RHS of Eq. \ref{eq:self}. To re-obtain this equation we multiply by $\xi_i^{\mu}$ and sum over $\mu$, using the definition of $L_i^{\mu}$ from Eqs. \ref{eq:lk},

\begin{eqnarray}
\sum_{\mu\neq 1}m_i^{\mu}\xi_i^{\mu} & = & G_i g_i\sum_{\mu\neq 1}\frac{\xi_i^{\mu}(1-a_i)}{C d_{\mu}} \left( c_{ii}+\sum_{j=1}^N c_{ji}\left\{ K_{ij}^{\mu}+\sum_{k=1}^N K_{ik}^{\mu}K_{kj}^{\mu}+\dots\right\}\right)+\nonumber\\
 + &\sum_{l\neq i}&G_l g_l \sum_{\mu\neq 1} \frac{\xi_i^{\mu}(\xi_l^{\mu}-a_l)}{C d_{\mu}} \left(c_{il}+\sum_{j=1}^N c_{jl}
\left\{ K_{ij}^{\mu}+\sum_{k=1}^N K_{ik}^{\mu}K_{kj}^{\mu}+\dots\right\} \right).\label{eq:selfnoise}
\end{eqnarray}

Let us first treat the first term of Eq. \ref{eq:selfnoise}, corresponding to $\gamma_i \sigma_i$ in Eq. \ref{eq:self}. Taking into account that $c_{ii}=0$ (no self-excitation), only the contribution including the curly brackets survives. As shown in \cite{yasser_2004}, each term inside the curly brackets, containing the product of multiple $K$'s, is different only to a vanishing order from the product of independent averages, each one corresponding to the sum of $K_{ab}$ over all pre-synaptic neurons $b$. In this way, 

\begin{eqnarray}
G_i g_i(1-a_i)\sum_{\mu\neq 1}\frac{\xi_i^{\mu}}{C d_{\mu}}\sum_{j,l_1\dots l_n=1}^N c_{ji} K_{il_1}^{\mu} \left[\prod_{o=1}^{n-2}K_{l_o l_{o+1}}^{\mu}\right]K_{l_nj}^{\mu}\simeq \nonumber \\
\simeq \alpha G_i g_i a_i(1-a_i)\frac{C}{N} \left<\frac{1}{d_{\mu}^{n+1}} \right>_{\mu}(a \Omega )^n,\label{eq:omega}
\end{eqnarray}
where we have introduced $\alpha\equiv p/C$, or the memory load normalized by the number of connections per neuron. The $\left<\dots \right>_{\mu}$ brackets symbolize an average over the index $\mu$ and $\Omega $ is a variable of order $1$ defined by

\begin{equation}
\Omega \equiv \frac{1}{a N}\sum_{j=1}^N a_j(1-a_j)G_j'g_j.\label{eq:om}
\end{equation}
Adding up all the terms with different powers of $\Omega$ in Eq. \ref{eq:omega} results in

\begin{equation}
\gamma_i \sigma_i=\alpha a_i(1-a_i) g_i \frac{C}{N} \left< \frac{\Omega}{d_{\mu} (d_{\mu}/a-\Omega )} \right>_{\mu}G_i\label{eq:gam}.
\end{equation}
Since $\Omega$ does not depend on $\mu$, if $d_{\mu}=a$ for all $\mu$ the average results simply in the classical $\Omega/(1-\Omega)$ factor.

As postulated in the ansatz, the second term in Eq. \ref{eq:selfnoise} is a sum of many independent contributions and can thus be thought of as a gaussian noise. Its mean is zero by virtue of the factor $(\xi_l^{\mu}-a_l)$, uncorrelated with both $\xi_i^\mu$ (by hypothesis) and $d_{\mu}$ (negligible correlation). Its variance is given by

\begin{equation}
\left< \left< \rho_i^2 \right> \right>=\left< \left< \sum_{l\neq i}G_l^2 g_l^2 \sum_{\mu\neq 1} \frac{\xi_i^{\mu}(\xi_l^{\mu}-a_l)^2}{C^2 d_{\mu}^2} \left(c_{il}+\sum_{j=1}^N c_{jl}\left\{ K_{ij}^{\mu}+\sum_{k=1}^N K_{ik}^{\mu}K_{kj}^{\mu}+\dots\right\} \right)^2\right> \right>
\end{equation}
which corresponds to the first and only surviving term of Eq. \ref{eq:variance4}, the other three terms vanishing for identical reasons. Distributing the square in the big parenthesis and repeating the steps of Eq. \ref{eq:omega} this results in

\begin{eqnarray}
\left< \left< \rho_i^2\right> \right>&=& \alpha a_i \left\{ \left<\frac{1}{d_{\mu}^2}\right>_{\mu}+2\frac{C}{N}\left<\frac{\Omega}{d_{\mu}^2(d_{\mu}/a-\Omega)}\right>_{\mu}+\frac{C}{N} \left< \frac{\Omega^2}{d_{\mu}^2(d_{\mu}/a-\Omega)^2}\right>_{\mu} \right\}\times \nonumber \\
& &\times \sum_{\mu \neq 1}\frac{1}{C}\sum_{l \neq i}(\xi_l^{\mu}-a_l)^2 g_l^2\frac{C}{N}G_l^2.
\end{eqnarray}

If we define
\begin{equation}
q\equiv \left\{\dots\right\}\frac{1}{N}\sum_{l=1}^N G_l^2 a_l(1-a_l) g_l^2\label{eq:q}
\end{equation}
including the whole content of the curly brackets from the previous equation, then the variance of the gaussian noise is simply $\alpha a_i q$, and the second term of Eq. \ref{eq:self} becomes
\begin{equation}
\rho_i z_i=\sqrt{\alpha a_i q} z_i
\end{equation}
with $z_i$, as before, an independent normally-distributed random variable with unitary variance. The initial hypothesis of Eq. \ref{eq:self} is, thus, self consistent.

Taking into account these two contributions, the mean field experienced by a neuron $i$ when retrieving pattern $1$ is

\begin{equation}
h_i=\xi_i^{1} m+\alpha a_i(1-a_i) G_i g_i \frac{C}{N} \left< \frac{\Omega}{d_{\mu}(d_{\mu}/a-\Omega)} \right>_{\mu}+\sqrt{\alpha q a_i} z_i, \label{eq:meanfield}
\end{equation}

where we have used $m_i^1\simeq m$ and

\begin{equation}
m\equiv \frac{1}{N d_1}\sum_{j=1}^N (\xi_j^1-a_j) g_j \sigma_j \label{eq:m}
\end{equation}
is a variable measuring the weighted overlap between the state of the network and the pattern $1$, which together with $q$ (Eq. \ref{eq:q}) and $\Omega$ (Eq. \ref{eq:om}) form the group of macroscopic variables describing the possible stable states of the system. While $m$ is a variable related to the signal that pushes the activity toward the attractor, $q$ and $\Omega$ are noise variables. Diluted connectivity is enough to make the contribution of $\Omega$ negligible (in which case the diluted equations \cite{hfsp} are re-obtained), while $q$ gives a relevant contribution as long as the memory load is significantly different from zero, $\alpha=p/C>0$.

To simplify the analysis we adopt the zero temperature limit ($\beta\rightarrow \infty $), which turns the sigmoidal function of Eq. \ref{eq:activation} into a step function. To obtain the mean activation value of neuron $i$, the field $h_i$ defined by Eq. \ref{eq:meanfield} must be inserted into Eq. \ref{eq:activation} and the equation in the variable $\sigma_i$ solved. This equation is

\begin{equation}
\sigma_i=\Theta \left[\xi_i^{1} m+\alpha a_i(1-a_i) \sigma_i g_i \frac{C}{N} \left< \frac{\Omega}{d_{\mu} (d_{\mu}/a-\Omega)} \right>_{\mu}+\sqrt{\alpha q a_i} z_i -U\right],\label{eq:theta}
\end{equation}
where $\Theta[x]$ is the Heaviside function yielding $1$ if $x>0$ and $0$ otherwise. When $z_i$ has a large enough modulus, its sign determines one of the possible solutions, $\sigma_i=1$ or $\sigma_i=0$. However, for a restricted range of values, $z_-\leq z_i\leq z_+$, both solutions are possible. Using the definition of $\gamma_i$ in Eq. \ref{eq:gam} to simplify notation, we can write $z_+=(U-\xi_i^{1} m)/\sqrt{\alpha q a_i}$ and $z_-=(U-\xi_i^{1} m- \gamma_i ))/\sqrt{\alpha q a_i}$. A sort of Maxwell rule must be applied to choose between the two possible solutions \cite{Shiino_1993}, by virtue of which the point of transition between the $\sigma_i=0$ and the $\sigma_i=1$ solutions is the average between the two extremes
\begin{equation}
y_{\xi}\equiv \frac{z_++z_-}{2}=\frac{U-\xi_i^{1} m-\gamma_i/2 }{\sqrt{\alpha q a_i}}\label{eq:y}.
\end{equation}

Inserting Eq. \ref{eq:theta} into Eq. \ref{eq:m} yields

\begin{equation}
m=\frac{1}{N d_1}\sum_{j=1}^N (\xi_j^1-a_j) g_j \int_{-\infty} ^{\infty } Dz \Theta \left[z-y_{\xi}\right],
\end{equation}
where we have introduced the average over the independent normal distribution $Dz$ for $z_j$. This expression can be integrated resulting in

\begin{equation}
m=\frac{1}{N d_1}\sum_{j=1}^N (\xi_j^1-a_j ) g_j \phi[y_{\xi}].\label{eq:m2},
\end{equation}
where we define
\begin{equation}
\phi(y_{\xi})\equiv \frac{1}{2}\left\{1+{\rm erf} \left[ \frac{y_{\xi}}{\sqrt{2}}\right] \right\}.
\end{equation}

Following the same procedure, Eq. \ref{eq:q} can be rewritten as

\begin{eqnarray}
q&=&
\left\{ \left<\frac{1}{d_{\mu}^2}\right>_{\mu}+2\frac{C}{N}\left<\frac{\Omega}{d_{\mu}^2(d_{\mu}/a-\Omega)}\right>_{\mu}+\frac{C}{N} \left< \frac{\Omega^2}{d_{\mu}^2(d_{\mu}/a-\Omega)^2}\right>_{\mu} \right\}\times \nonumber\\
& &\times \frac{1}{N}\sum_{j=1}^N a_j(1-a_j) g_j^2 \phi(y_{\xi}).\label{eq:q2}
\end{eqnarray}

Before repeating these steps for the variable $\Omega$ we note that

\begin{equation}
\int Dz G_j'= \frac{1}{\sqrt{\alpha q a_j}}\int Dz \frac{\partial \sigma_j}{\partial z}=\frac{1}{\sqrt{\alpha q a_j}}\int Dz z \sigma_j,
\end{equation}
where we have applied integration by parts. Eq. \ref{eq:omega} results then in

\begin{equation}
\Omega=\frac{1}{N a}\sum_{j=1}^N \frac{a_j(1-a_j ) g_j}{\sqrt{2\pi\alpha q a_j}} \exp\left\{-\frac{y_{\xi}^2}{2}\right\}.\label{eq:w2}
\end{equation}

\begin{figure}[ht]
\centerline{\hbox{\epsfig{figure=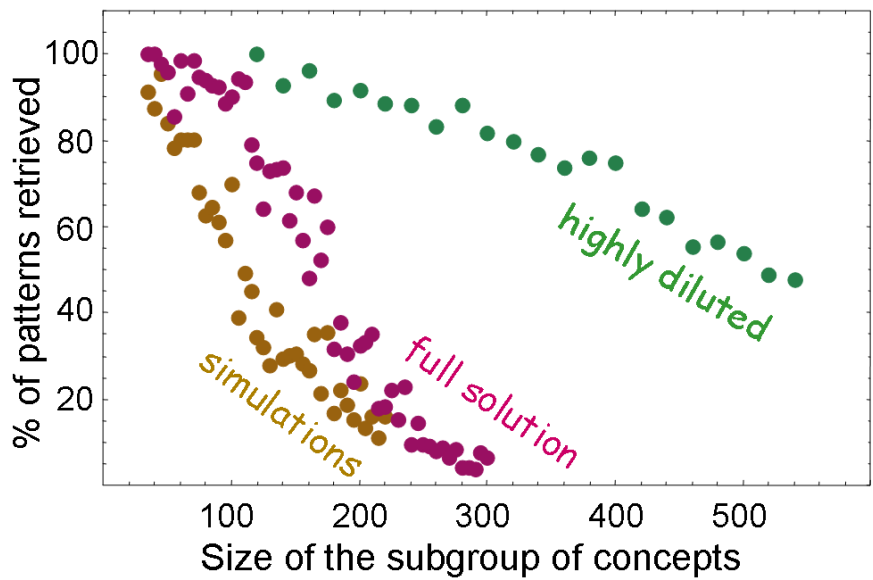,width=12cm,angle=0}}}
\caption{Simulations and numerical solutions of the equations of a network storing random subgroups of patterns taken from the feature norms of McRae and colleagues. The performance of the network depends strongly on the size of the subgroup. Though this is observed in the highly diluted approximation, the decay in performance is not enough to explain the data. It is the full solution with $g(x)=1$ that results in a good fit of the simulations. In each simulation, the number of neurons equals the number of features describing some of the stored concepts, and there is full connectivity between neurons, $C=N$.}
\label{fig:1}
\end{figure}

Eqs. \ref{eq:m2}, \ref{eq:q2} and \ref{eq:w2} define the stable states of the network. Retrieval is successful if the stable value of $m$ is close to $1$. In Figure \ref{fig:1} we show the performance of a fully connected network storing the feature norms of McRae and colleagues \cite{McRae_norms} in three situations: theoretical prediction for a diluted network as in \cite{hfsp}, theoretical prediction for a fully connected network calculated from Eqs. \ref{eq:m2}-\ref{eq:w2} and the actual simulations of the network. The figure shows that the fully connected theory better approximates the simulations, performed with random subgroups of patterns of varying size $p$ and full connectivity for each neuron, $C=N$, equal to the total number of features involved in the representation of the subgroup of concepts.

Finally, we can rewrite Eqs. \ref{eq:m2}-\ref{eq:w2} in a continuous way by introducing two types of popularity distribution across neurons:

\begin{equation}
F(x) = P(a_i=x)
\end{equation}
as the global distribution, and
\begin{equation}
f(x) = P(a_i=x|\xi_i^{1}=1)
\end{equation}
as the distribution related to the pattern that is being retrieved. 

The equations describing the stable values of the variables become

\begin{eqnarray}
m&=&\int_0^1 f(x) g(x) (1-x)\phi(y_1)-\frac{1}{d_1}\int_0^1 \left[F(x)-d_1 f(x)\right] g(x) x \phi(y_0) \nonumber\\
q&=&
\left\{ \left<\frac{1}{d_{\mu}^2}\right>_{\mu}+2\frac{C}{N}\left<\frac{\Omega}{d_{\mu}^2(d_{\mu}/a-\Omega)}\right>_{\mu}+\frac{C}{N} \left< \frac{\Omega^2}{d_{\mu}^2(d_{\mu}/a-\Omega)^2}\right>_{\mu} \right\}\times \nonumber\\
& & \times \left\{ d_1\int_0^1 f(x) g^2(x) x(1-x) \phi(y_1)+\int_0^1 \left[F(x) -d_1 f(x)\right]g^2(x) x(1-x)\phi(y_0) \right\}\label{eq:sys}\nonumber\\
\Omega&=&\frac{d_1}{a}\int_0^1 f(x) g(x) \frac{x(1-x)}{\sqrt{2\pi\alpha q x}}\exp(-y_1^2/2)+\frac{1}{a}\int_0^1 [F(x)-d_1 f(x)] g(x) \frac{x(1-x)}{\sqrt{2\pi\alpha q x}}\exp(-y_0^2/2)\label{eq:sys2},
\end{eqnarray} 
where, adapted from Eq. \ref{eq:y}, 
\begin{equation}
y_{\xi}=\frac{1}{\sqrt{\alpha q x}}\left(U-\xi m -\alpha x(1-x)g(x)\frac{C}{2N}\left<\frac{\Omega}{d_{\mu} ( d_{\mu}/a-\Omega )} \right>_{\mu} \right).
\end{equation}

\section{The storage of feature norms}
\label{norms}

In \cite{hfsp} we have shown that the \textit{robustness} of a memory in a highly diluted network is inversely related to the \textit{information} it carries. More specifically, a stored memory needs a minimum number of connections per neuron $C_{min}$ that is proportional to 

\begin{equation}
I_f\equiv \int_0^1 f(x)x(1-x) dx.\label{eq:info}
\end{equation}
In this way, if connections are randomly damaged in a network, the most informative memories are selectively lost.

\begin{figure}[ht]
\centerline{\hbox{\epsfig{figure=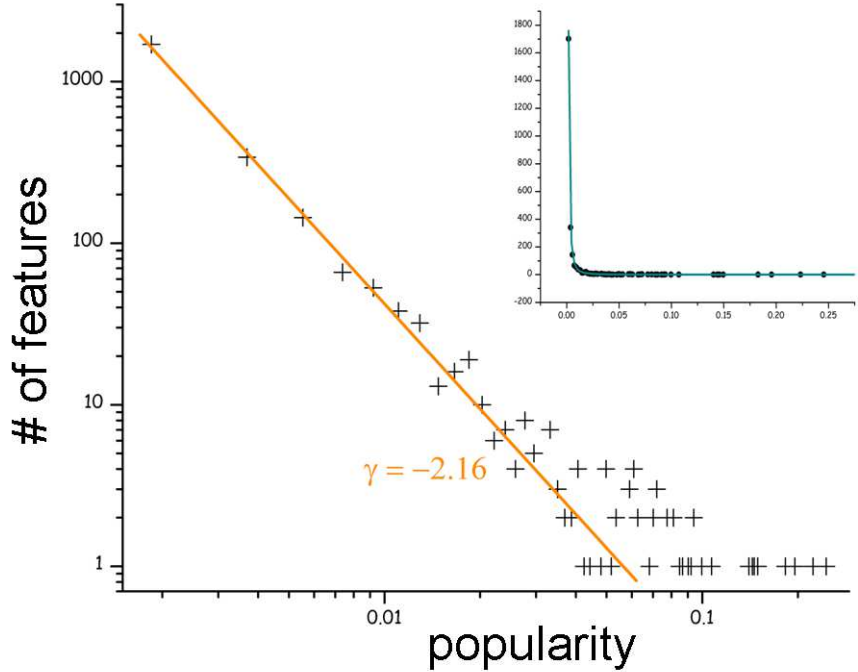,width=12cm,angle=0}}}
\caption{The popularity distribution $F(x)$ of the feature norms is a power law, with $\gamma \simeq 2.16$. Note that both axes are logarithmic. In the inset, the same plot appears with linear axes, including the corresponding fit.}
\label{fig:2}
\end{figure}
The distribution $F(x)$ affects the retrievability of all memories. As we have shown in the same paper, it is typically a function with a maximum near $x=0$. The relevant characteristic of $F(x)$ is its tail for large $x$. If $F(x)$ decays fast enough, the minimal connectivity scales like

\begin{equation}
C_{min}\propto p I_f \log \left[\frac{I_F}{aI_f}\right],
\end{equation}
where $I_F$ corresponds to the same pseudo-information function as in Eq. \ref{eq:info}, but using the distribution $F(x)$. If $F(x)$ decays exponentially ($F(x)\sim \exp(-x/a)$), the scaling of the minimal connectivity is the same, with only a different logarithmic correction,

\begin{equation}
C_{min}\propto p I_f \log^2 \left[\frac{I_F}{aI_f}\right].
\end{equation}
The big difference appears when $F(x)$ has a tail that decays as slow as a power law ($F(x)\sim x^{-\gamma}$). The minimal connectivity is then much larger

\begin{equation}
C_{min}\propto \frac{p I_f}{a} \log \left[\frac{a^{\gamma-2}}{I_f}\right]
\end{equation}
since the sparseness, measuring the global activity of the network, is in cortical networks $a\ll 1$. Unfortunately, as can be seen in Figure \ref{fig:2}, the distribution of popularity $F(x)$ for the feature norms of McRae and colleagues is of this last type. This is the reason why, as shown in Figure \ref{fig:1}, the performance of the network is very poor in storing and retrieving patterns taken from this dataset. In a fully connected network as the one shown in the figure, a stored pattern can be retrieved as long as its minimal connectivity $C_{min}\leq N$, the number of connections per neuron. Along the $x$ axis of the Figure, representing the number of patterns from the norms stored in the network, the average of $I_f$ is rather constant, $p$ and $N$ increase proportionally and $a$ decreases, eventually taking $C_{min}$ over the full connectivity limit. 

In the following subsections, we analyze different ways to increase this poor storage capacity and effectively store and retrieve the feature norms in an autoassociative memory.

\subsection{Adding uninformative neurons}

\begin{figure}[ht]
\centerline{\hbox{\epsfig{figure=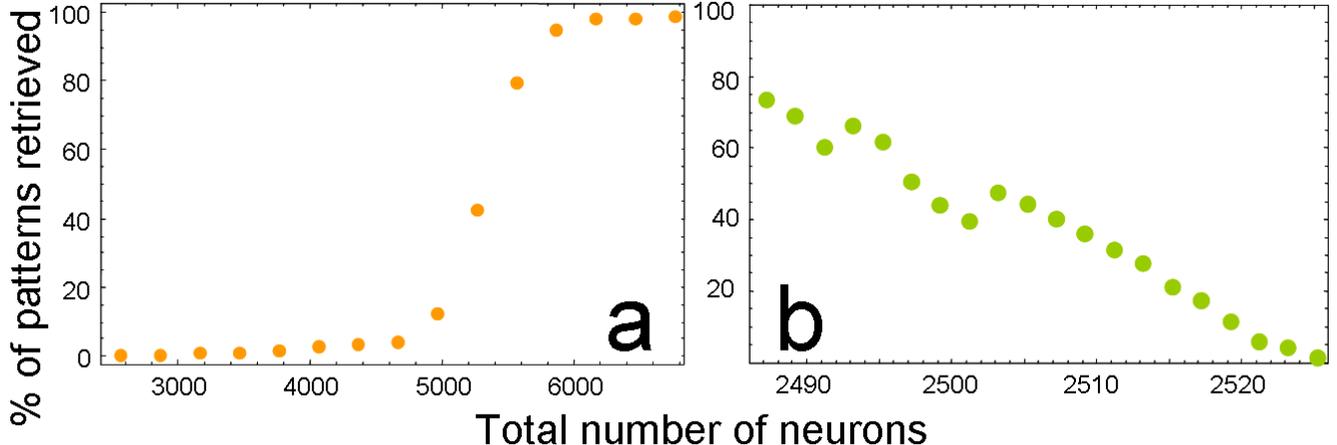,width=18cm,angle=0}}}
\caption{Adding or taking neurons affects the overall distribution $F(x)$ and, thus, the performance of the network. The starting point for both situations is $2526$ neurons corresponding to all the features in the norms. {\bf a}: Adding 5 neurons with minimal popularity per pattern is enough to get $100\%$ performance. Note that the transition is sharp. {\bf b}: Removing the $40$ most informative neurons also results in an improved performance, in this case of $80\%$ of the stored patterns.}
\label{fig:3}
\end{figure}

As discussed in \cite{hfsp}, a way to increase the storage capacity of the network in general terms is to push the distribution $F(x)$ toward the smaller values of $x$. One possibility is to add neurons with low information value (i.e. with low popularity) so as to make $I_f$ smaller in average without affecting the sparseness $a$ too much. In Figure \ref{fig:3}a we show that the full set of patterns from the feature norms can be stored and retrieved if $5$ new neurons per pattern are added, active in that particular pattern and in no other one. 

\subsection{Removing informative neurons}

A similar effect on the distribution $F(x)$ can be obtained by eliminating selectively the most informative neurons. In Figure \ref{fig:3}b we show that if the full set of patterns is stored a retrieval performance of $\sim 80\%$ is achieved if the $40$ more informative features are eliminated. We estimate that $100\%$ performance should be achieved if around $60$ neurons were selectively eliminated.

It is not common in the neural literature to find a poor performance that is improved by damaging the network. This must be interpreted in the following way. The connectivity of the network is not enough to sustain the retrieval of the stored patterns, too informative to be stable states of the system. By throwing away information, the system can be brought back to work. However, a price is being payed: the representations are impoverished since they no longer contain the most informative features.

\subsection{Popularity-modulated weights}

A final way to push the distribution $F(x)$ toward low values of $x$ can be figured from Eqs. \ref{eq:sys2}. Indeed, $g(x)$ can be thought of as a modulator of the distributions $F(x)$ and $f(x)$. Inspired in \cite{hfsp}, if $g(x)$ decays exponentially or faster, the storage capacity of a set of patterns with any decaying $F(x)$ distribution should be brought back to a $C_{min}\propto pI_f$ dependence, without the $a^{-1}\gg 1$ factor. 

\begin{figure}[ht]
\centerline{\hbox{\epsfig{figure=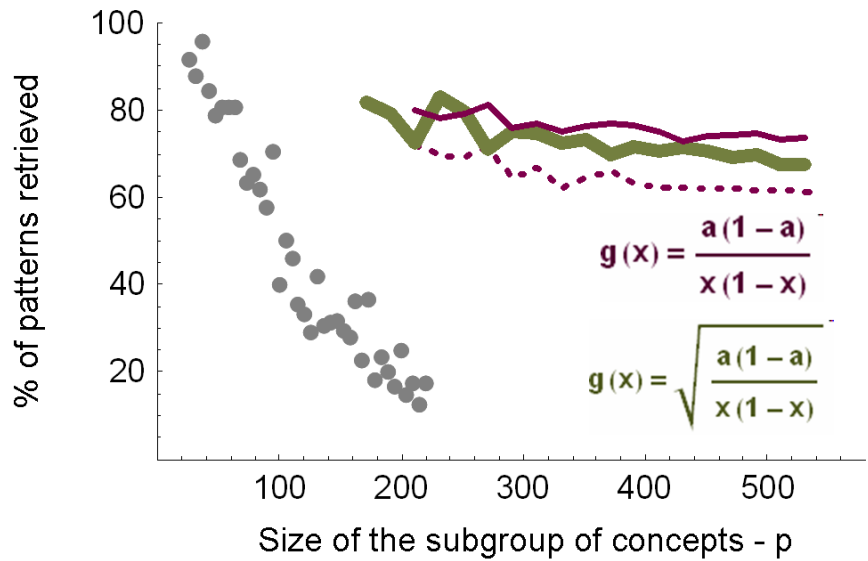,width=12cm,angle=0}}}
\caption{Simulations (dashed line) and theoretical predictions (solid line) of a network storing subgroups of patterns of varying size taken from McRae and colleagues feature norms with a popularity-modulated hebbian learning rule. The thin violet lines use a value of $g(x)$ inversely proportional to $x(1-x)$, normalized so as to maintain the average field of order $1$. The thick green line corresponds to a $g(x)$ inversely proportional to $\sqrt{x(1-x)}$. Following our predictions, the exact form of $g(x)$ does not affect the general performance, which is substantialy improved with respect to the simulations with $g(x)=1$, copied from Figure \ref{fig:2} in grey dots.}
\label{fig:4}
\end{figure}

In Figure \ref{fig:4} we analyze two possible $g(x)$ functions that favor low over high values of $x$:
\begin{eqnarray}
g_1(x)&=&\frac{a(1-a)}{x(1-x)}\\
g_2(x)&=&\sqrt{\frac{a(1-a)}{x(1-x)}}.
\end{eqnarray}

The storage capacity of the network increases drastically in both cases. Furthermore, we estimate that $\sim 60\%$ of the lost memories in the figure suffer from a too high value of the threshold $U$, set, as in all simulations in the paper, to $0.6$. This value was chosen to maximize the performance in the previous simulations. However, with a much more controlled noise, the optimal threshold should be lower, generally around $m/2$. Setting the threshold in this level could maybe improve even further the performance of the network. 

\section{Discussion}
We have presented the full non diluted solution describing the stable states of a network that stores correlated patterns. A simple Hebbian learning rule is applicable as long as neurons can be treated as statistically independent. In order to analyze the storage of the patterns taken from the feature norms of McRae and colleagues, we include in the learning rule the possibility that the global activity is different for each pattern. The full solution explains the poor performance of autoassociative networks storing the feature norms \cite{McRae_1997, McRae_1999, McRae_2006}. We show that this data has a popularity distribution decaying as a power law, the worse of the cases analyzed in \cite{hfsp}.

The three proposed solutions aiming to improve the storage capacity of the network have a very different scope. Adding unpopular neurons is a feasible solution for McRae and colleagues. In the procedure of collecting the norms, a threshold is used to decide whethter or not a given feature is relevant enough to be included in the dataset. Lowering the threshold would result in a set of patterns with many more very uninformative features. In second place, the elimination of very informative neurons in a damaged network could be achieved by damaging selectively the most active ones, bringing back the network to work. Finally, the modulation of  synaptic strength following pre-synaptic popularity can be considered to be an intermediate solution between the two extremes. Whether or not it is a cortical strategy applied to deal  with correlated representations is a question for which we have yet no experimental evidence.

\bibliographystyle{apalike}
\bibliography{correlations}

\end{document}